\documentclass[%
 reprint,
 amsmath,amssymb,
 aps,
prb,
 superscriptaddress
]{revtex4-1}

\usepackage{graphicx}
\usepackage{dcolumn}
\usepackage{bm}
\usepackage{braket}

\begin{document}

\preprint{APS/123-QED}

\title{Nodal-line semimetal superlattices}

\author{Kazuki Yokomizo}
\affiliation{Department of Physics, Tokyo Institute of Technology, 2-12-1 Ookayama, Meguro-ku, Tokyo, 152-8551, Japan}
\author{Hiroaki Yamada}
\affiliation{Department of Physics, Tokyo Institute of Technology, 2-12-1 Ookayama, Meguro-ku, Tokyo, 152-8551, Japan}
\author{Shuichi Murakami}
\affiliation{Department of Physics, Tokyo Institute of Technology, 2-12-1 Ookayama, Meguro-ku, Tokyo, 152-8551, Japan}
\affiliation{TIES, Tokyo Institute of Technology, 2-12-1 Ookayama, Meguro-ku, Tokyo, 152-8551, Japan}%




%
\begin{abstract}
Spatial modulations, such as superlattices, to realize topological materials have recently been studied in theoretical and experimental works. In this paper, we investigate properties of the superlattices of the nodal-line semimetal (NLS) and the normal insulator. We consider two types of superlattices, with the stacking direction being perpendicular or parallel to the plane where the nodal line lies. In particular, we show that when the stacking direction is parallel to the plane, the nodal lines remain but they change their shapes because of the folding of the Brillouin zone. We also study the superlattices with magnetization. One can expect that the quantum anomalous Hall (QAH) phase emerges in some cases, depending on the direction of the magnetization. If the magnetization is along the $C_2$-invariant axis, the superlattice becomes the Weyl semimetal phase if the $C_2$-invariant axis intersects the nodal lines, and otherwise it becomes the QAH phases.
%
\end{abstract}
\pacs{Valid PACS appear here}
\maketitle
%
%
\section{\label{sec1}Introduction}

Recently, topology has been shown to play a crucial role in condensed matter physics \cite{Yan2012,Ando2013,Chiu2016}. Realizations of topological phases in condensed matter physics have started from the discovery of the quantum Hall effect \cite{Klitzing1980}. After that, various topologically-insulating systems, such as the quantum anomalous Hall (QAH) systems \cite{Haldane1988,Onoda2008,Liu2008,Yu2010} and the topological insulators \cite{Kane2005,Bernevig2006,Fu2007}, have been theoretically proposed and experimentally observed over the past decade.

In addition to these topologically-insulating phases, topological semimetals have been also introduced. These materials have been classified into various categories. One example is a Weyl semimetal (WSM) \cite{Murakami2007,Wan2011} with point nodes at which the valence and the conduction bands touch. Another example is a nodal-line semimetal (NLS), which has line-shaped degeneracies between the valence and conduction bands in the bulk, and it has been intensively studied \cite{Burkov2011nls,Fang2015,Heikkila2015,Hyart2016,Li2017,Yu2018}. In recent years, various NLSs have been theoretically predicted \cite{Xie2015,Chan2016,Zeng2015,Kim2015,Yu2015,Bian2016,Liang2016,Xu2017,Hirayama2017,Huang2016,Mook2017,Tateishi2018,Wang2018,Li2018,Zhang2018} and experimentally confirmed \cite{Schoop2016,Yamakage2016,Bian2016ex,Takane2018,Liu2018,Sato2018,Wang2018-ex}.

On the other hand, spatial modulations, such as superlattices, to realize topological materials have recently been studied in theoretical \cite{Bernevig2006,Liu2008,Burkov2011,Takahashi2011,Jiang2012,Xu2015,Yokomizo2017,Lau2017,Behrends2017,Gong2018} and experimental \cite{Konig2007,Gaoyuan2018} works. One of the pioneering works is on a superlattice of a topological insulator (TI) and a normal insulator (NI) with magnetization; it was theoretically proposed that the superlattice can realize a WSM phase \cite{Burkov2011}. Another example is a superlattice of a WSM and a NI, giving rise to QAH phases \cite{Yokomizo2017}. In these proposals, spatial modulations are useful means to manipulate bulk-band structure, and in particular to realize topological band structures by manipulating singularities in the momentum space. Meanwhile, superlattices of a NLS and a NI has not been studied thus far, to the authors' knowledge.

In this paper, we study superlattices of a NLS and a NI. Here, we consider the class of the spinless NLSs with inversion symmetry (IS) and time-reversal symmetry (TRS), and thereby the nodal lines are characterized by the $\pi$ Berry phase. We investigate the superlattices with the stacking direction being either perpendicular or parallel to the plane where the nodal line lies, and we call them pattern A and pattern B, respectively. We find that the resulting phase diagrams include only the NLS phase and the NI phase. Furthermore, in pattern B with magnetization added, the phase diagram shows rich physics, including the QAH phases with various values of the Chern number, similarly to the WSM-NI superlattice \cite{Yokomizo2017}. To be more specific, when the system has $C_2$ symmetry and there are no intersections between the $C_2$-invariant axis and the nodal lines in the NLS superlattice, the QAH phase appears when the magnetization is added along the $C_2$-invariant axis.

This paper is organized as follows. In Sec.~\ref{sec2}, we investigate the properties of NLS superlattices with patterns A and B from the effective model of the NLS. In Sec.~\ref{sec3}, we also study how the QAH phases appear in the superlattices with magnetization. Furthermore, we calculate the band structure using the lattice model and compare the results of the effective model with those of the lattice model in Sec.~\ref{sec4}. Finally, we summarize and discuss the results in Sec.~\ref{sec5}. Throughout the paper, we restrict ourselves to the cases with the IS and assume that the spin-orbit couping is negligible.
\section{\label{sec2}Superlattice from the effective model}
\subsection{\label{subsec2-1}Effective model of the NLS}

\begin{figure}[]
\begin{tabular}{cc}
\begin{minipage}[]{0.5\hsize}
\begin{center}
\includegraphics[width=4.5cm,height=4.0cm]{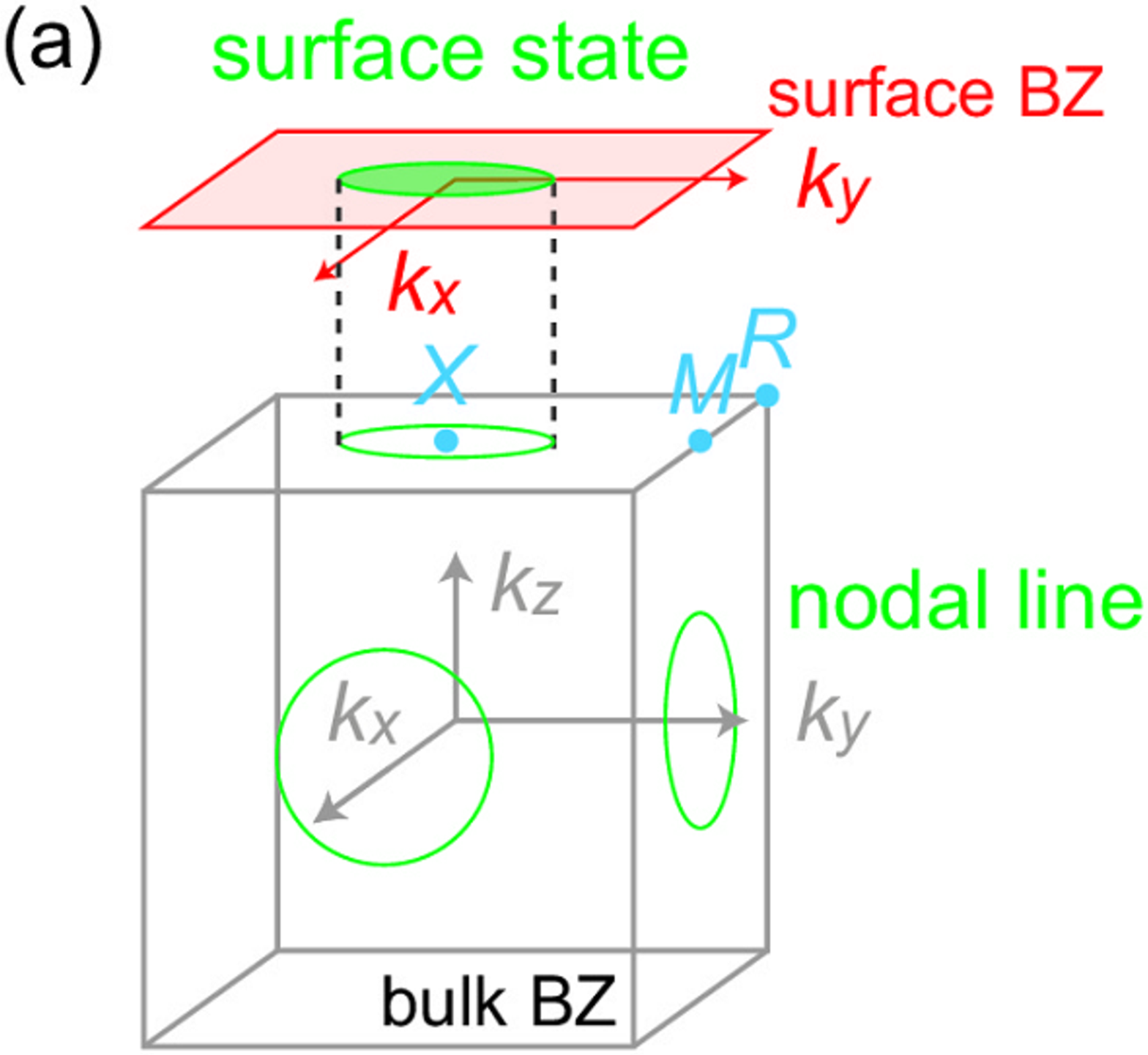}
\end{center}
\end{minipage}
\begin{minipage}[]{0.5\hsize}
\begin{center}
\includegraphics[width=3.4cm,height=4.0cm]{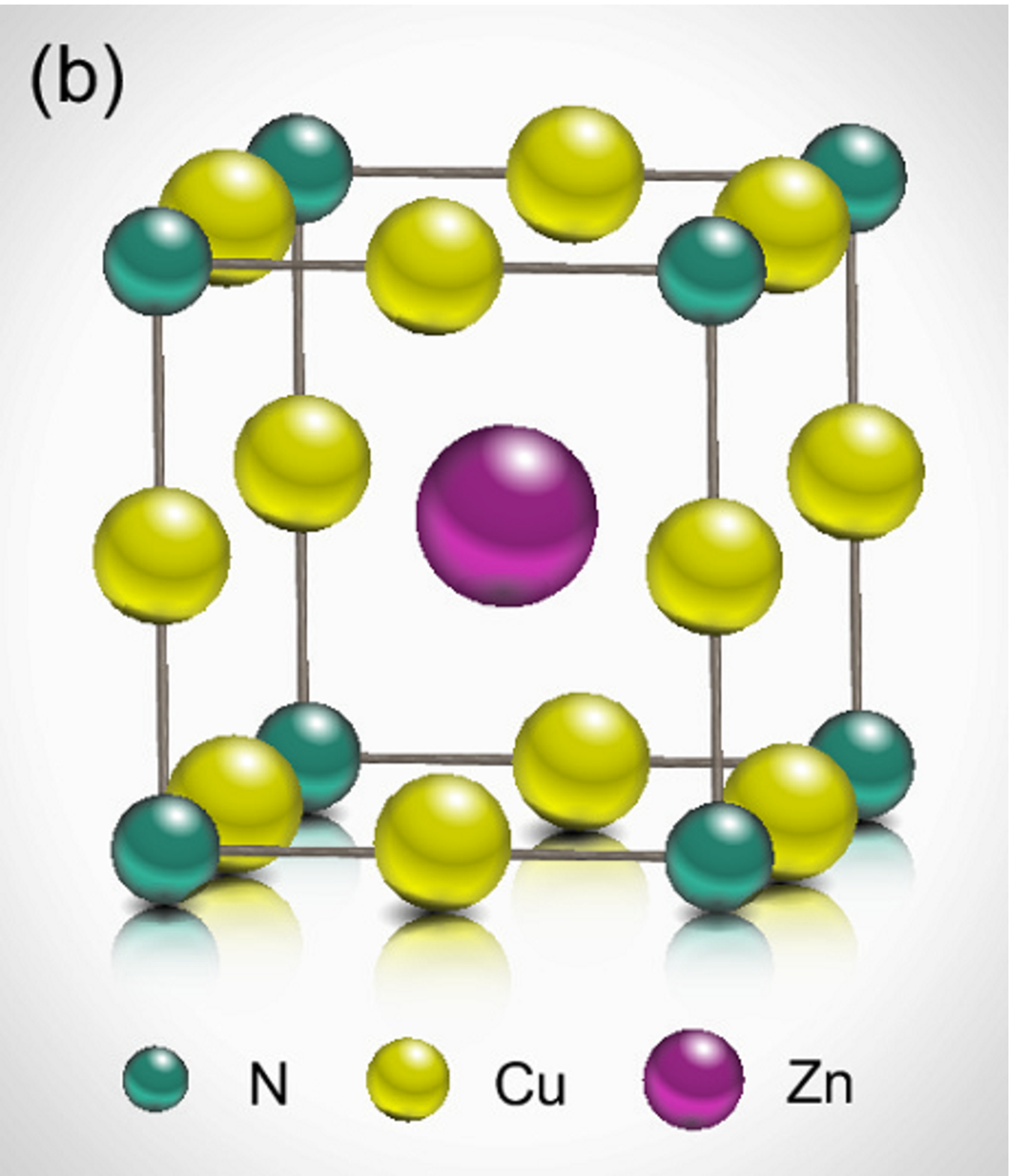}
\end{center}
\end{minipage}
\end{tabular}
\caption{\label{fig1}(Color online) (a) Bulk and surface Brillouin zones (BZs) of ${\rm Cu_3NZn}$. There are nodal lines near the $X$ points represented by the green lines. When the nodal line is projected to the surface Brillouin zone, the drumhead surface states appear in the region surrounded by the nodal line. (b) Crystal structure of ${\rm Cu_3NZn}$.}
\end{figure}

Here, we review an effective model of the NLS proposed in Ref.~\onlinecite{Kim2015}. It is showed that ${\rm Cu_3NZn}$ has nodal lines around the $X$ points [Fig.~\ref{fig1} (a)]. The crystal structure of ${\rm Cu_3NZn}$ is a cubic anti-${\rm ReO_3}$ structure, intercalated with ${\rm Zn}$ atoms at the body center of the cubic unit cell of ${\rm Cu_3N}$ as shown in Fig.~\ref{fig1} (b). Let $d$ be the lattice constant. Then, the Hamiltonian around the $X$ point ${\bm X}^{z}=\pi\hat{\bm z}/d$ can be expanded as
\begin{equation}
H_X=v_zk_z\tau_y+\left(\Delta\varepsilon+g_\perp k_\perp^2+g_zk_z^2\right)\tau_z,
\label{eq1}
\end{equation}

\noindent where $k_\perp$ is given by $k_\perp^2=k_x^2+k_y^2$, $v_z$, $\Delta\varepsilon$, $g_\perp$, and $g_z$ are constants, and $v_z$, $g_\perp$, $g_z$ are positive. Here, we have imposed both the TRS represented by $T=K$, with $K$ being the complex conjugation operator, and the IS represented by $P=\tau_z$, together with the $D_{4h}$ point group symmetries at $X$. The Pauli matrices $\tau_i\hspace{3pt}\left(i=1,2,3\right)$ act on the space spanned by the $A_{1g}$ and $A_{2u}$ states. The energy eigenvalues are given by
\begin{equation}
E_X=\pm\sqrt{v_z^2k_z^2+\left(\Delta\varepsilon+g_\perp k_\perp^2+g_zk_z^2\right)^2}.
\label{eq2}
\end{equation}

\noindent Henceforth, we set the Fermi energy to be $E_F=0$. If $\Delta\varepsilon<0$, the valence and conduction bands are degenerate at $k_\perp^2=-\Delta\varepsilon/g_\perp$ and $k_z=0$, and the degeneracy forms the nodal line with the radius $\sqrt{-\Delta\varepsilon/g_\perp}$ at the Fermi energy. We note that since the nodal lines in ${\rm Cu_3NZn}$ have a non-trivial $Z_2$ invariant defined from the parity eigenvalues \cite{Kim2015} and a mirror symmetry with respect to the $k_z=0$ plane, the nodal lines are preserved both by topology and by the mirror symmetry. On the other hand, if $\Delta\varepsilon>0$, the system is in the NI phase.

In the following, we consider two patterns of superlattices shown in Fig.~\ref{fig2} (a) and Fig.~\ref{fig3} (a), separately.

\subsection{\label{subsec2-2}Superlattice: Pattern A}

\begin{figure}[]
\centering
\includegraphics[width=8.3cm,height=7.25cm]{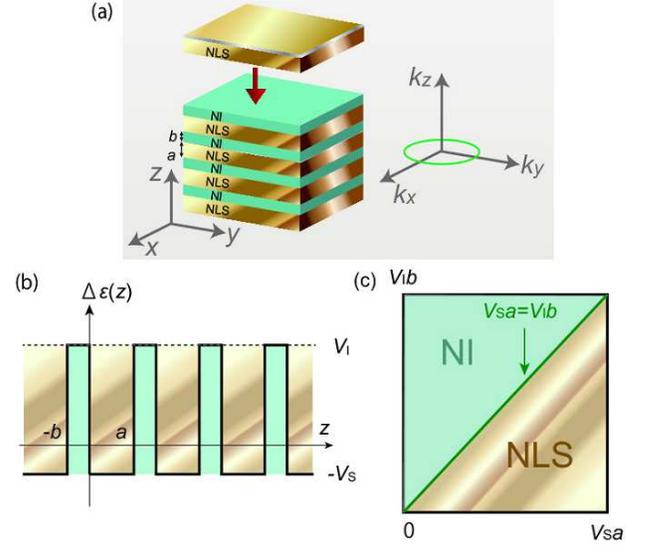}
\caption{\label{fig2} (a) Schematic figure of the superlattice and the direction of the nodal line with pattern A. (b) Spatial dependence of $\Delta\varepsilon\left(z\right)$. $\Delta\varepsilon\left(z\right)=-V_S$ in the NLS layers and $\Delta\varepsilon\left(z\right)=V_I$ in the NI layers. (c) The phase diagram of the NLS superlattice with pattern A.}
\end{figure}

We study a superlattice of a NLS and a NI stacked perpendicularly to the nodal-line plane as shown in Fig.~\ref{fig2} (a). We call it pattern A. To realize this superlattice, we periodically modulate $\Delta\varepsilon$ in the effective Hamiltonian Eq.~({\ref{eq1}}) between positive and negative values as shown in Fig.~\ref{fig2} (b). Thus, the Hamiltonian of the NLS superlattice can be written as
\begin{eqnarray}
\begin{array}{l}
H=-iv_z\partial_z\tau_y+\left[\Delta\varepsilon\left(z\right)+g_\perp k^2_\perp\right]\tau_z, \vspace{3pt}\\
\Delta\varepsilon\left(z\right)=\left\{ \begin{array}{ll}
-V_S  & \left(0\leq z\leq a\right), \vspace{5pt}\\
V_I   & \left(-b\leq z\leq0\right),
\end{array}\right. \vspace{3pt}\\
\Delta\varepsilon\left(z+\left(a+b\right)\right)=\Delta\varepsilon\left(z\right),
\end{array}
\label{eq3}
\end{eqnarray}

\noindent where $k_\perp=\sqrt{k_x^2+k_y^2}$. Here, $\Delta\varepsilon\left(z\right)=-V_S<0$ and $\Delta\varepsilon\left(z\right)=V_I>0$ represent the NLS and the NI layers, respectively.

One can solve the eigenvalue problem of this model similarly to the Kr\"{o}nig-Penny model \cite{Kronig1931,Yokomizo2017}. After straightforward but lengthy calculation, we find that the nodal line in the superlattice appears only when $V_Sa>V_Ib$, and it is located at
\begin{equation}
k_x^2+k_y^2=\frac{V_Sa-V_Ib}{g_\perp\left(a+b\right)},~~k_z=0,
\label{eq4}
\end{equation}

\noindent where $k_z$ is the Bloch wave number along the $z$ direction. Thus, the superlattice is in the NLS phase only when $V_Sa>V_Ib$, and the nodal line disappears at $V_Sa=V_Ib$. When $V_Sa<V_Ib$, the superlattice is in the NI phase.

In Eq.~(\ref{eq4}), we consider how the nodal line changes across the phase transition between the NLS and the NI phases. When $b=0$, the superlattice corresponds to the original NLS, and the nodal line with the radius  $\sqrt{V_S/g_\perp}$ is located on the $k_z=0$ plane. By increasing $b$, the radius of the nodal line gradually decreases to $\displaystyle\sqrt{\frac{V_Sa-V_Ib}{g_\perp\left(a+b\right)}}$, and the nodal line eventually shrinks to a point at ${\bm k}={\bm 0}$ when $b=\left(V_S/V_I\right)/a$. In the case $b>\left(V_S/V_I\right)/a$, the nodal line disappears, and the superlattice becomes the NI. As a result, the phase diagram is shown in Fig.~\ref{fig2} (c).
\subsection{\label{subsec2-3}Superlattice: Pattern B}

\begin{figure}[b]
\centering
\includegraphics[width=6.0cm,height=6.55cm]{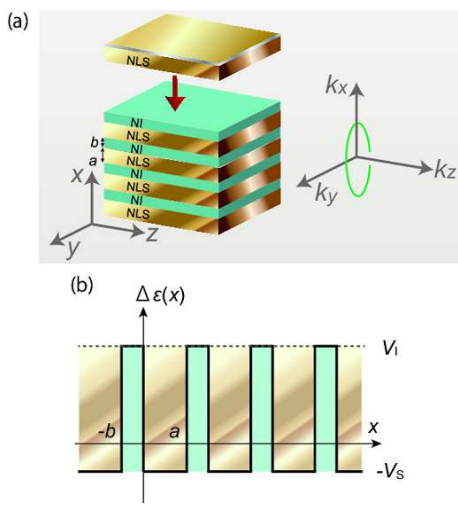}
\caption{\label{fig3} (a) Schematic figure of the superlattice and the direction of the nodal line plane with pattern B. The nodal line in the NLS lies within the $k_z=0$ plane. (b) Spatial dependence of $\Delta\varepsilon\left(x\right)$.}
\end{figure}

In this section, we consider a NLS-NI superlattice with the stacking direction being parallel to the nodal-line plane as shown in Fig.~\ref{fig3} (a). Similarly to Sec.~\ref{subsec2-2}, the Hamiltonian describing the superlattice is given by
\begin{eqnarray}
\begin{array}{l}
H=v_zk_z\tau_y+\left[\Delta\varepsilon\left(x\right)+g_\perp\left(-\partial_x^2+k^2_y\right)\right]\tau_z, \vspace{3pt}\\
\Delta\varepsilon\left(x\right)=\left\{ \begin{array}{ll}
-V_S  & \left(0\leq x\leq a\right), \vspace{5pt}\\
V_I   & \left(-b\leq x\leq0\right),
\end{array}\right. \vspace{3pt}\\
\Delta\varepsilon\left(x+\left(a+b\right)\right)=\Delta\varepsilon\left(x\right),
\end{array}
\label{eq5}
\end{eqnarray}

\noindent where $\Delta\varepsilon\left(x\right)$ is the same as that in Sec.~\ref{subsec2-2}, and the schematic figures of the superlattice and spatial dependence of $\Delta\varepsilon\left(x\right)$ are shown in Fig.~\ref{fig3}.

We can calculate the eigenstates and the energies in the similar way to as in the previous subsection. Since the Hamiltonian Eq.~(\ref{eq5}) has the mirror symmetry with respect to the $k_z=0$ plane, we expect that the nodal lines in the superlattice lie within the plane. Therefore, we can obtain the equation for the position of the nodal lines:
\begin{eqnarray}
&&\frac{V_I^\prime-V_S^\prime+2k_y^2}{2\sqrt{V_S^\prime-k_y^2}\sqrt{V_I^\prime+k_y^2}}\sin\sqrt{V_S^\prime-k_y^2}a\sinh\sqrt{V_I^\prime+k_y^2}b \nonumber\\
&&+\cos\sqrt{V_S^\prime-k_y^2}a\cosh\sqrt{V_I^\prime+k_y^2}b=\cos k_x\left(a+b\right),
\label{eq6}
\end{eqnarray}

\noindent where $k_x$ is the Bloch wave number along the $x$ direction, $V_S^\prime=V_S/g_\perp$, and $V_I^\prime=V_I/g_\perp$. When $b$ is sufficiently large, the superlattice is in the NI phase, while by increasing $a$, the superlattice becomes the NLS phase. Here, since the original nodal lines are preserved both by the TRS and by the mirror symmetry \cite{Kim2015}, once the nodal lines appear in the superlattice, they cannot disappear due to the $\pi$ Berry phase guaranteed by symmetries. Therefore, the topologically-insulating phases cannot appear in the superlattice by a small perturbation preserving the symmetries.

\begin{figure*}[]
\centering
\includegraphics[width=15.0cm,height=4.8cm]{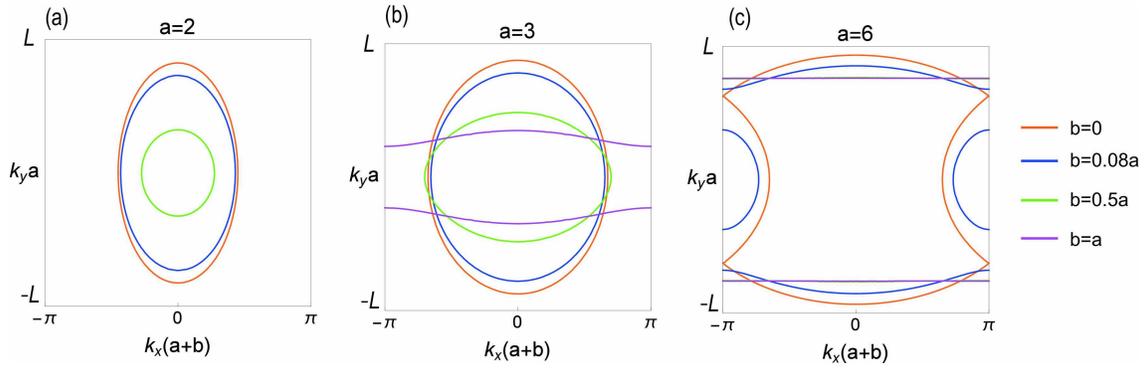}
\caption{\label{fig4}(Color online) Nodal lines in the superlattice for (a) $a=2$, (b) $a=3$, and (c) $a=6$. Values of the parameter are fixed as $V_S=0.5$, $V_I=1$, and $b$ is varied as $b=0$, $b=0.08a$, $b=0.5a$, and $b=a$. $L$ is defined as $L=\sqrt{V_S}a$. We note that in (a), nodal lines do not exist when $b=a$, and in (c), the green ($b=0.5a$) and purple ($b=a$) nodal lines almost overlap.}
\end{figure*}

We investigate how the nodal lines appear in the superlattice from Eq.~(\ref{eq6}). The nodal lines for various values of $a$ and $b$ are shown in Fig.~\ref{fig4}. When $a$ is small as shown in Fig.~\ref{fig4} (a), by increasing $b$, the nodal line shrinks. It does not cross the boundary of the Brillouin zone. On the other hand, when $a$ is sufficiently large as shown in Figs.~\ref{fig4} (b) and (c), the nodal line crosses the boundary. When $a=3$, the nodal lines recombine and gradually becomes straight along the $k_x$ axis. Furthermore, when $a=6$, the nodal lines recombine to form three line nodes. Then, one shrinks while the others gradually become straight and parallel to the $k_x$ axis. To understand the origin of these almost straight nodal lines, we consider the $b\rightarrow\infty$ limit which corresponds to the quantum well of the NLS. In this limit, Eq.~(\ref{eq6}) does not depend on the Bloch wave number $k_x$ as shown in Appendix \ref{sec6}, and the gapless states exist in the quantum well when $a$ is above some critical values. Therefore, these states corresponds to the nodal lines parallel to the $k_x$ direction.

\begin{figure}[]
\begin{tabular}{cc}
\begin{minipage}[]{0.5\hsize}
\begin{center}
\includegraphics[width=4.2cm,height=3.5cm]{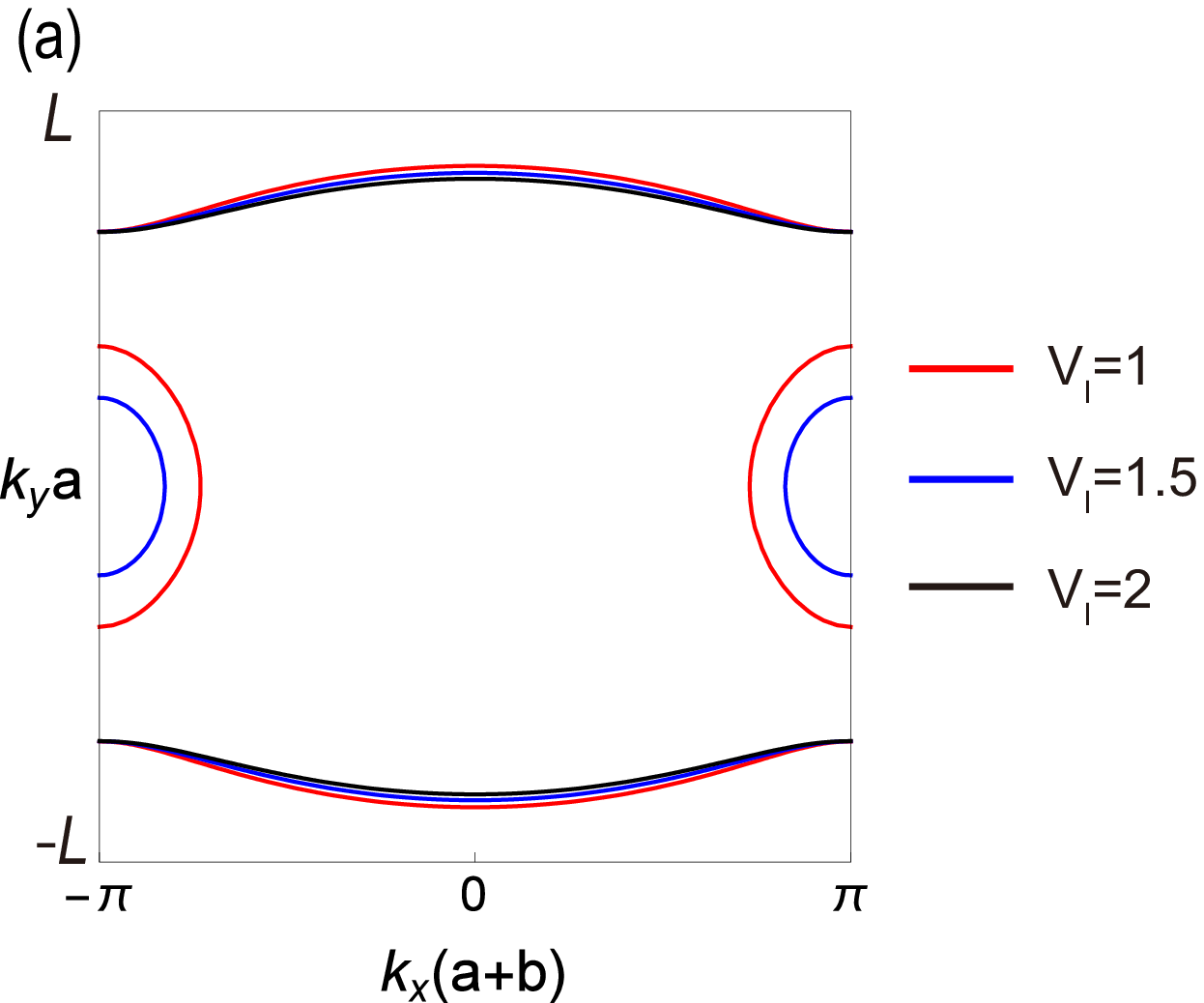}
\end{center}
\end{minipage}
\begin{minipage}[]{0.5\hsize}
\begin{center}
\includegraphics[width=4.2cm,height=3.5cm]{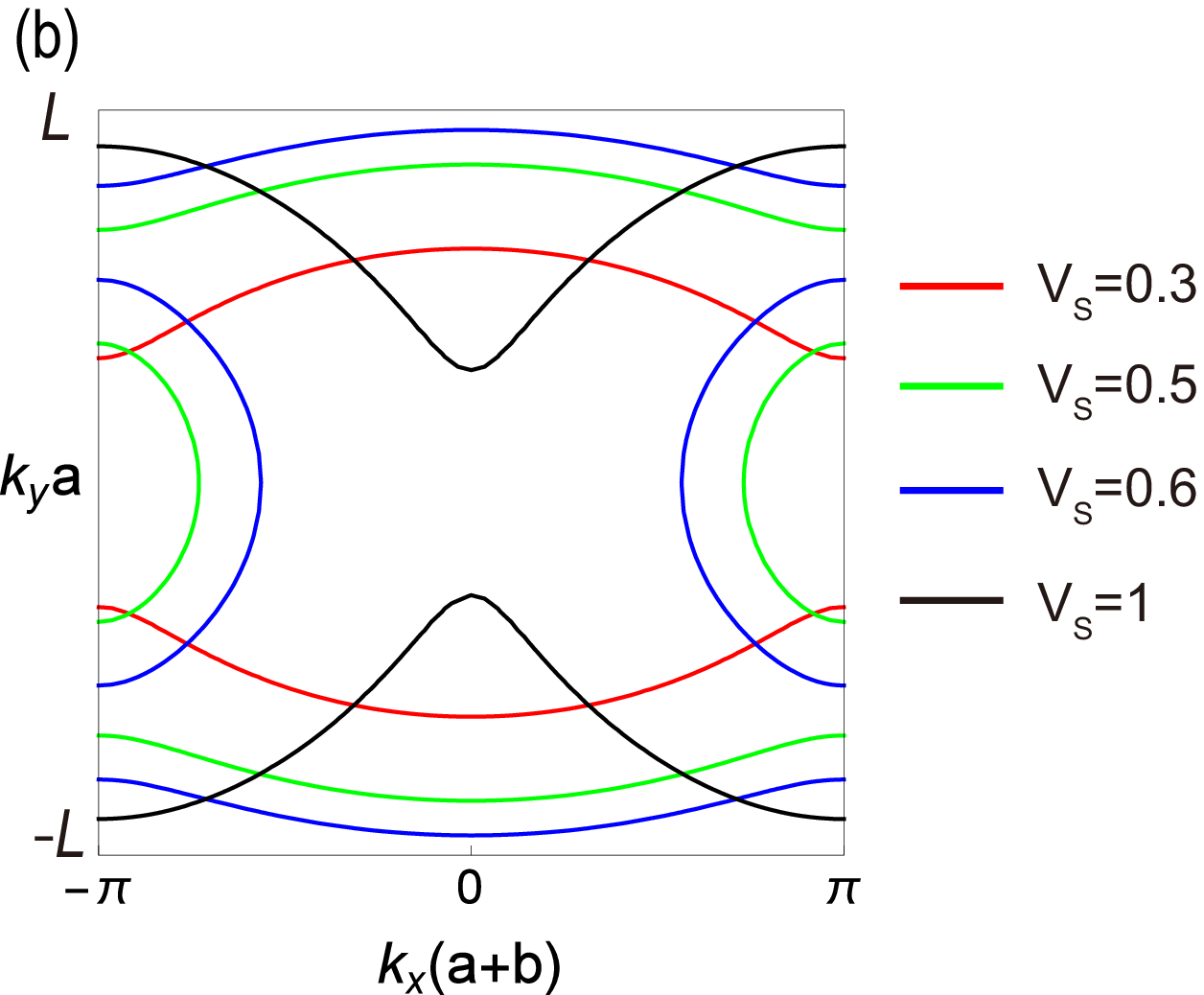}
\end{center}
\end{minipage}
\end{tabular}
\caption{\label{fig5}(Color online) Nodal lines in the superlattice with various values of the parameters $V_I$ and $V_S$. The thickness of each layer is set as $a=6$, $b=0.08a$. In (a), $V_S$ is fixed to be $V_S=0.5$, and $V_I$ is varied as $V_I=1$, $V_I=1.5$, and $V_I=2$. In (b), $V_I$ is fixed to be $V_I=1$, and $V_S$ is varied as $V_S=0.3$, $V_S=0.5$, $V_S=0.6$, and $V_S=1$. $L$ is defined as $L=\sqrt{V_S}a$. We note that in (a), almost straight nodal lines around $k_ya\simeq\pm0.7a$ for the three cases $V_I=1$, $V_I=1.5$, and $V_I=2$ almost overlap.}
\end{figure}

The circular nodal line shrinks not only by increasing $b$ but also by increasing the parameter $V_I$. In the effective Hamiltonian Eq.~(\ref{eq1}), $V_I$ represents the bulk gap of the NI. When $a=6$, $b=0.08a$, and $V_I=1$, among the nodal lines, the circular one crosses the Brillouin zone boundary, and the others are almost parallel to the $k_x$ axis as shown in the red lines [Fig.~\ref{fig5} (a)]. Then, by increasing $V_I$, the circular nodal line gradually shrinks (blue lines, $V_I=1.5$), and eventually disappears (black line, $V_I=2$).

On the other hand, by increasing the parameter $V_S$ which represents the radius of the original nodal line, the almost straight nodal line disappears. This can be seen in Fig.~\ref{fig5} (b). When $a=6$, $b=0.08a$, and $V_S=0.3$, the nodal lines appear as shown by the red lines. By increasing $V_S$, the nodal lines which are almost parallel to the $k_x$ axis gradually move along the $k_y$ axis, and finally disappear at the Brillouin zone boundary. On the other hand, the circular nodal line also extends, and eventually recombines to form new nodal lines shown by the black lines.

\section{\label{sec3}Superlattice with magnetization}

In this section, we investigate what happens to the NLS superlattices if the magnetization appears. It has been shown that spinless NLSs undergo a transition to spinless WSMs by breaking the TRS under some conditions \cite{Okugawa2017}. Therefore, we can expect that the NLS superlattice with magnetization may behave similarly to the WSM superlattice in the previous paper \cite{Yokomizo2017}.

In the NLS superlattice with pattern A, we can immediately find that only the NI and the WSM phases appear when the magnetization is introduced. On the other hand, in the NLS superlattice with pattern B, we expect that the QAH phases may emerge, depending on the direction of the magnetization. We classify the pattern B into two cases; we call the case of the magnetization ${\bm m}$ parallel to the stacking direction ${\bm n}$ pattern B-1 and that of the magnetization ${\bm m}$  perpendicular to the stacking direction ${\bm n}$ pattern B-2.
\subsection{\label{subsec3-1}Superlattice: Pattern B-1}

We study the NLS superlattice, with the magnetization introduced parallel to the stacking direction. For this purpose, we introduce the perturbation $V_T=mk_y\tau_x$ to the effective Hamiltonian Eq.~(\ref{eq1}). The effective model without the perturbation $V_T$ has the TRS and $D_{4h}$ symmetry, and when we introduce the perturbation $V_T$, the IS and the $C_2$ symmetry around the magnetization direction remain. From this symmetry argument, this perturbation represents the magnetization along the $x$ axis, and the parameter $m$ represents the magnitude of the magnetization.

We can solve the energy eigenvalue problem with the perturbation, similarly to Sec.~\ref{subsec2-3}. By using the fact that the Weyl nodes appear along the stacking direction because of the symmetries \cite{Yokomizo2017}, we conclude that the band gap closes at $k_y=k_z=0,E=0$ and
\begin{eqnarray}
&&\frac{V_I-V_S}{2\sqrt{V_SV_I}}\sin\sqrt{V_S/g_\perp}a\sinh\sqrt{V_I/g_\perp}b \nonumber\\
&&+\cos\sqrt{V_S/g_\perp}a\cosh\sqrt{V_I/g_\perp}b=\cos k_x\left(a+b\right),
\label{eq7}
\end{eqnarray}

\noindent If Eq.~(\ref{eq7}) has real solutions for $k_x$, the superlattice is in the WSM phase, and otherwise it is in a bulk-insulating phase. We note that whether the superlattice becomes the WSM phase or a bulk-insulating phase does not depend on the value of the magnetization $m\left(\neq0\right)$ but on the parameters $V_S$ and $V_I$.

\begin{figure}[]
\centering
\includegraphics[width=6.5cm,height=6.6cm]{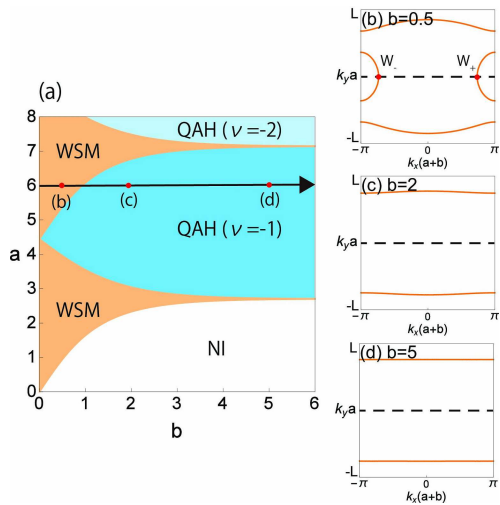}
\caption{\label{fig6}(Color online) Phase diagram of the superlattice with pattern B-1. We set $V_S=0.5$ and $V_I=1.0$. $\nu$ is the Chern number of the QAH phases. (b)-(d) Nodal lines in the original NLS superlattice shown as orange lines. The thickness of the NLS layer is fixed to be $a=6$, while the thickness of the NI layer is taken as $b=0.5$ in (b), $b=2.0$ in (c), and $b=5.0$ in (d). The dashed black lines represent the $C_2$-invariant axis parallel to the magnetization. When the magnetization is added, in (b), two Weyl nodes $\left({\rm W}_\pm\right)$ appear at the intersections between the nodal line and the $C_2$-invariant axis, while in (c) and (d), Weyl nodes do not appear. Here, $L=\sqrt{V_S}a$. The change from (b) to (c) and (d) corresponds to the change of $b$ along the solid black line in (a).}
\end{figure}

We show the phase diagram of the superlattice in Fig.~\ref{fig6} (a) by changing $a$ and $b$. To see the physical origin for this phase diagram, we study how the phase changes along the $a=6$ line shown as the arrow in Fig.~\ref{fig6} (a). In Fig.~\ref{fig6} (b)-(d),  we show the positions of the nodal lines in the original NLS superlattice without the magnetization with $b=0.5$, $2$, and $5$ calculated from Eq.~(\ref{eq6}). In Ref.~\onlinecite{Okugawa2017}, it is shown that when the magnetization along the $C_2$-invariant axis is added to the NLS, the system becomes a WSM with the Weyl nodes appearing at intersections between the $C_2$-invariant axis and the nodal lines. It is indeed the case here. First of all, when $a=6$, $b=0.5$ and $m=0$, the nodal lines consist of the circular one and the two almost straight ones [Fig.~\ref{fig6} (b)]. Because the circular nodal line intersects with the $C_2$-invariant axis, it is natural that the superlattice is in the WSM phase [Fig.~\ref{fig6} (a)]. On the other hand, by increasing $b$, the circular nodal lines at $m=0$ disappear and the others eventually become almost straight as shown in Figs.~\ref{fig6} (c) and (d). Because there is no intersection between the $C_2$-invariant axis and nodal line, by adding the magnetization, no Weyl nodes appear. Then, the superlattice becomes the QAH phase [Fig.~\ref{fig6} (a)], which is analogous to the QAH phase in the NI-WSM superlattice in Ref.~\onlinecite{Yokomizo2017}. As $b$ is increased, the pair of the Weyl nodes created at ${\bm k}={\bm0}$ are annihilated pairwise at the boundary of the Brillouin zone. Therefore, we conclude that the QAH phase has the Chern number $\nu=-1$.

\begin{figure}[]
\centering
\includegraphics[width=8.5cm,height=4.5cm]{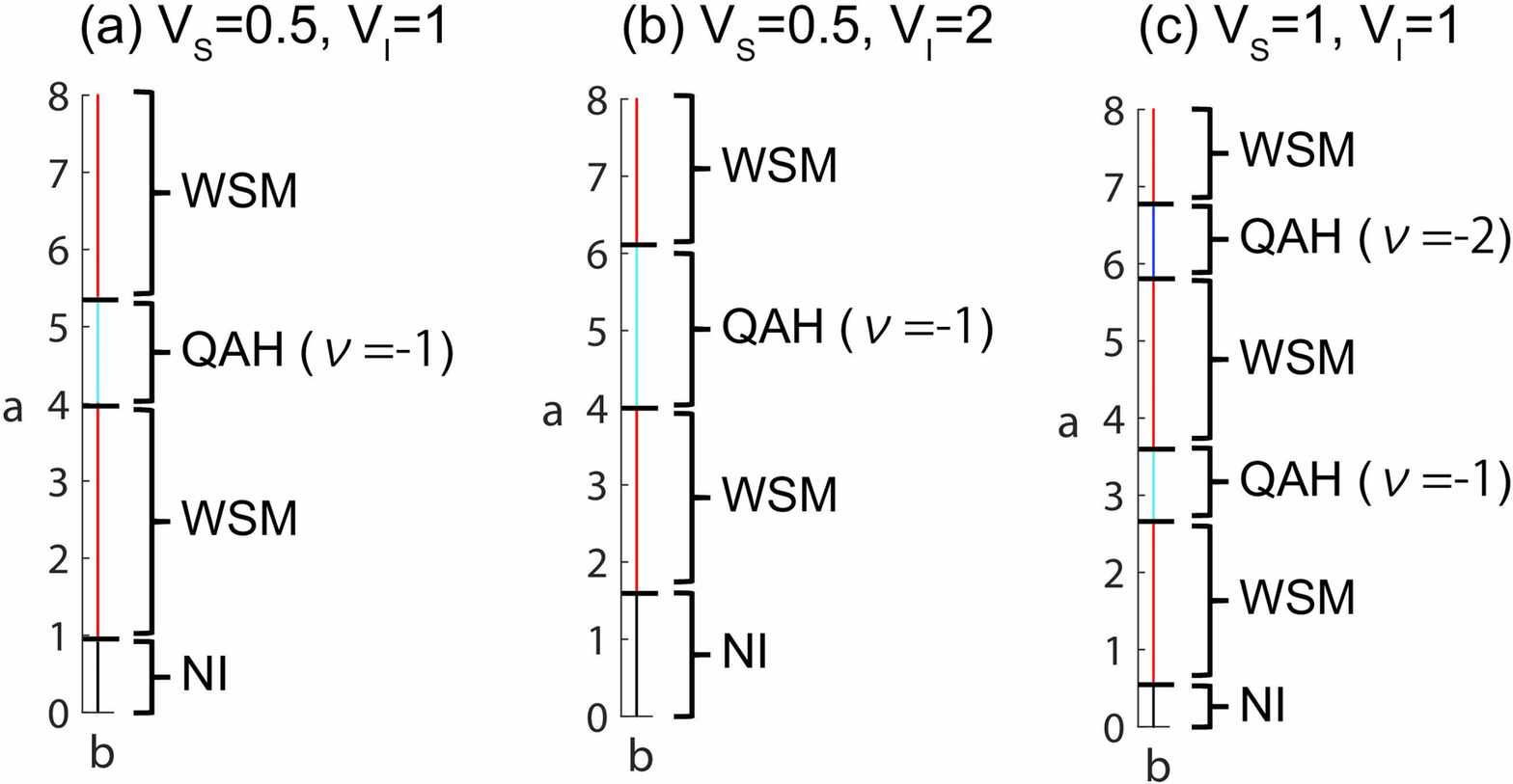}
\caption{\label{fig7} Phase diagrams of the superlattice with pattern B-1 with $b=0.48$ fixed and the values of other parameters being (a) $V_S=0.5$, $V_I=1$, (b) $V_S=0.5$, $V_I=2$, and (c) $V_S=1$, and $V_I=1$. When $a=6$, for example, by increasing $V_I$, the superlattice becomes the QAH phase with the Chern number $\nu=-1$ shown in (b) from the WSM phase shown in (a). On the other hand, by increasing $V_S$ with $a=6$ fixed, the WSM phase shown in (a) changes to the QAH phase with the Chern number $\nu=-2$ shown in (c).}
\end{figure}

Next, we study the phase transitions for $m\neq0$ by increasing the values of the parameters $V_S$ or $V_I$. For example, we fix $a=6$ and $b=0.48$. When we increase the value of $V_I$, a phase transition occurs from the WSM phase [Fig.~\ref{fig7} (a)] to the QAH phase with the Chern number $-1$ [Fig.~\ref{fig7} (b)]. Meanwhile, when we increase $V_S$, it gives rise to a transition from the WSM phase[Fig.~\ref{fig7} (a)] to the QAH phase with the Chern number $-2$ [Fig.~\ref{fig7} (c)]. Therefore, the transition from the WSM phase to the bulk-insulating phase occurs not only by increasing $b$ but also either by increasing the gap $V_I$ of the NI layer or increasing the radius of the original nodal line $V_S$.

\subsection{\label{subsec3-2}Superlattice: Pattern B-2}

In this subsection, we introduce the magnetization perpendicular to the stacking direction, i.e. the $y$ axis, of the superlattice. It is added to the Hamiltonian Eq.~(\ref{eq5}) in the form $V_T^\prime=mk_x\tau_x$.

\begin{figure}[]
\centering
\includegraphics[width=8.6cm,height=4.8cm]{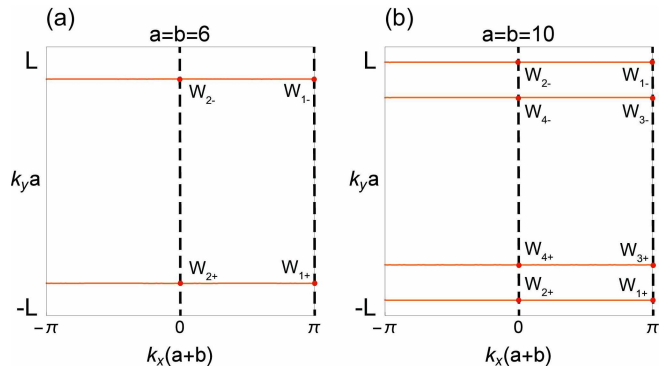}
\caption{\label{fig8}(Color online) Emergence of the Weyl nodes in the NLS superlattice with the magnetization. The orange lines represent the almost straight nodal lines in the original NLS superlattice with $m=0$ and the black dashed lines represent the $C_2$-invariant axis in the direction of the magnetization. The values of the parameters are set as $V_S=0.5$, $V_I=1$, $L=\sqrt{V_S}a$ and (a) $a=b=6$, (b) $a=b=10$.}
\end{figure}

In this case, the WSM phase appears in the superlattice if the $C_2$-invariant axis intersects the nodal lines, and otherwise the NI phase appears. In particular, when $a$ and $b$ are sufficiently large, the number of pairs of Weyl nodes increases since the number of the almost straight nodal lines also increase [Fig.~\ref{fig8}]. This change of the number of the Weyl nodes is not seen in the model of the NI-WSM superlattice in Ref.~\onlinecite{Yokomizo2017}. We note that topologically-insulating phases cannot appear since the $C_2$-invariant axis always intersects the almost straight nodal lines.

\section{\label{sec4}Superlattice from the lattice model}

In this section, we numerically investigate the behavior of the NLS-NI superlattices with a lattice model introduced in Sec.~\ref{subsec4-1}. We calculate the band structure of the superlattices with patterns A and B in Secs.~\ref{subsec4-2} and Secs.~\ref{subsec4-3}, respectively. Finally, we study the effect of magnetization in the NLS superlattice with pattern B in Sec.~\ref{subsec4-4}.
\subsection{\label{subsec4-1}NLS from the lattice model}

We construct a tight-binding model describing a NLS. For this purpose, we start from the continuum Hamiltonian Eq.~(\ref{eq1}). In a simple cubic lattice consisting of an $s$-like orbital and a $p_z$-like orbital per one site, we can construct the lattice Hamiltonian as
\begin{equation}
H_{\bm k}=\sum_ia_i\left({\bm k}\right)\sigma_i,
\label{eq8}
\end{equation}

\noindent where
\begin{eqnarray}
a_1\left({\bm k}\right)&=&0, \nonumber\\
a_2\left({\bm k}\right)&=&\frac{v}{d}\sin k_zd, \\
a_3\left({\bm k}\right)&=&\Delta\varepsilon+\frac{2g_\perp}{d^2}\left(2-\cos k_xd-\cos k_yd\right) \nonumber\\
&&+\frac{2g_z}{d^2}\left(1-\cos k_zd\right), \nonumber
\label{eq9}
\end{eqnarray}

\noindent and $d$ is the lattice constant. Here, we have $\Delta\varepsilon=-V_S$ for the NLS phase, while $\Delta\varepsilon=V_I$ for the NI phase.

In the following sections, we study a lattice model of the NI-NLS superlattice by modulating $\Delta\varepsilon$. We set $N_S$ and $N_I$ to be the numbers of the atomic layers within the NLS layer and the NI layer, respectively. Then, the thickness of the NLS layer and that of the NI layer are given by $a=N_Sd$ and $b=N_Id$, respectively.
\subsection{\label{subsec4-2}Superlattice: Pattern A}

\begin{figure}[]
\centering
\includegraphics[width=9.0cm,height=3.8cm]{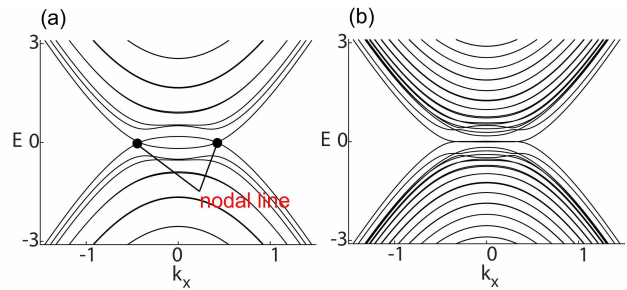}
\caption{\label{fig9} Band structure of the superlattices with pattern A for $k_y=k_z=0$. We set the parameters with $V_S=0.5$, $V_I=0.1$, $d=2$, $g_\perp=2$, $g_z$=1.2, and $v=1$. The thickness of the layers are (b) $N_S=10$, $N_I=5$ and (b) $N_S=10$, $N_I=30$. In (a), the superlattice becomes the NLS. We show the intersections between the nodal line and the $k_x$ axis. On the other hand, in (b), the superlattice does not have nodal lines and there is a tiny gap fo the entire values of $k_x$.}
\end{figure}

In this subsection, we study the NLS superlattice with pattern A, which has the stacking direction perpendicular to the nodal-line plane. We numerically calculate the band structure with the parameters $V_S=-0.5$ and $V_I=0.1$ as shown in Fig.~\ref{fig9}. We find that the superlattice has the nodal line in the bulk when $N_S=10$ and $N_I=5$ [Fig.~\ref{fig9} (a)]. By comparing this result with that of the continuum model, we find that this superlattice corresponds to the case of $V_Sa>V_Ib$ in Fig.~\ref{fig2} (c). As we have seen in Sec.~\ref{subsec2-2}, by increasing $b$, the nodal line shrinks and disappears in the superlattice. It is also the case for the lattice model as shown for $N_S=10$ and $N_I=30$ in Fig.~\ref{fig9} (b), where the superlattice is the NI phase. There is a tiny gap throughout the whole Brillouin zone.

Here, an almost flat band around $E=0$ is seen in the bulk as shown in Fig.~\ref{fig9} (b). We have discussed the appearance of this flat band in Ref.~\onlinecite{Yokomizo2017}. This behavior is caused by the drumhead surface states which appear on the interfaces between the NLS layers and the NI layers. Namely, since the adjacent drumhead interface states are separated by the insulating layers, these interface states hybridize to form the states near the Fermi energy in Fig.~\ref{fig9} (b), with small hybridization between them.

\subsection{\label{subsec4-3}Superlattice: Pattern B}

\begin{figure*}[]
\centering
\includegraphics[width=13.0cm,height=13.3cm]{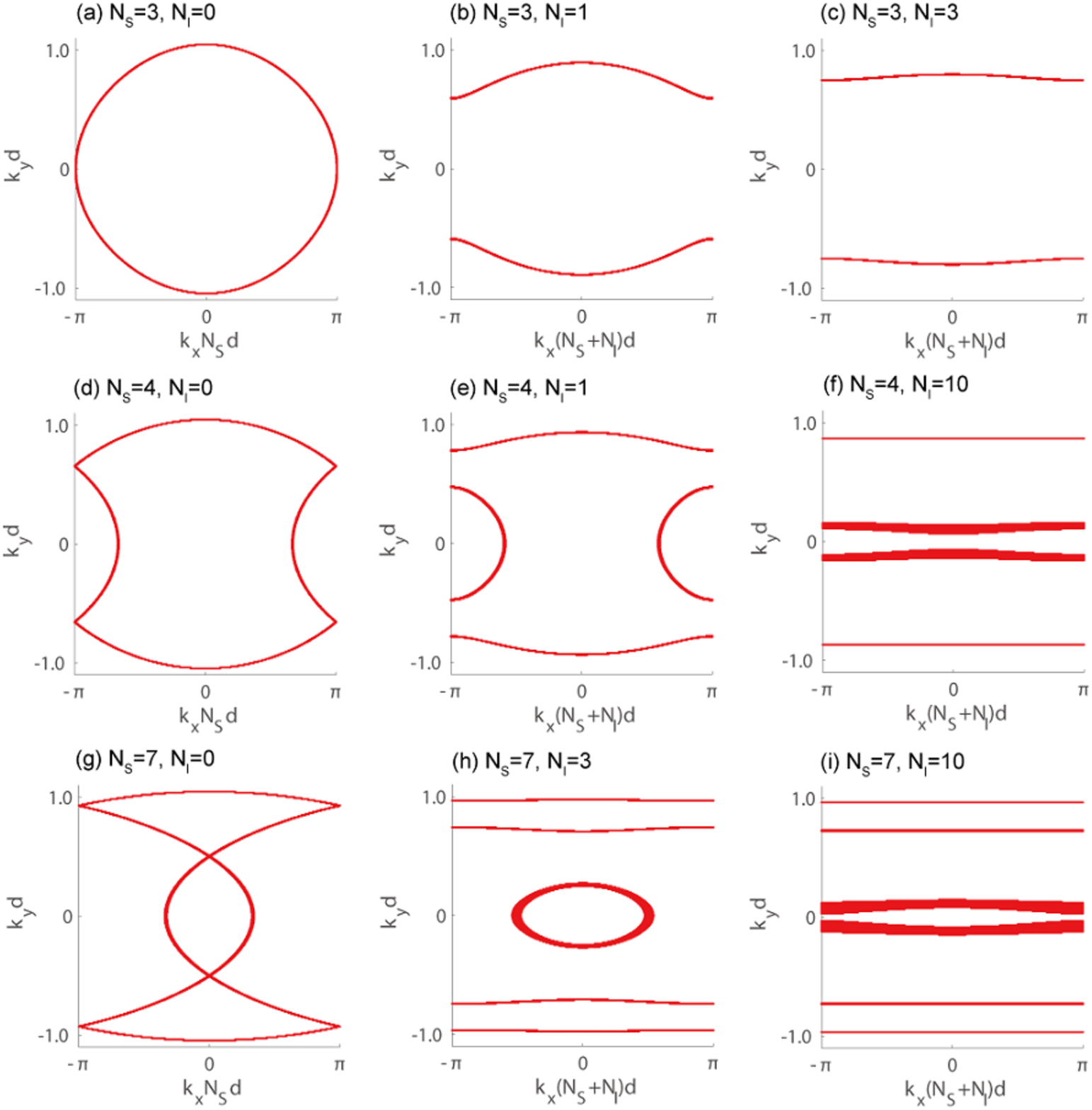}
\caption{\label{fig10} Nodal lines in the bulk and the superlattices on the $k_z=0$ plane. Here, to compare with the nodal lines of the NLS superlattice, we show the nodal lines of the bulk NLS by setting $N_I=0$, and in (a) $N_S=3$, (d) $N_S=4$, and (g) $N_S=7$. The nodal lines in the superlattices are shown for (b) $N_S=3$, $N_I=1$, (c) $N_S=3$, $N_I=3$, (e) $N_S=4$, $N_I=1$, (f) $N_S=4$, $N_I=10$, (h) $N_S=7$, $N_I=3$, and (i) $N_S=7$, $N_I=10$. Here, we set the parameters with $V_S=0.5$, $V_I=0.1$, $d=2$, $g_\perp=2$, $g_z$=1.2, and $v=1$. }
\end{figure*}

In this subsection, we calculate the band structure of the superlattices with pattern B as shown in Fig.~\ref{fig10}. We show the nodal lines around $E=0$ for various numbers of the NLS layers and of the NI layers. For comparison, we also show the cases with $N_I=0$ in Figs.~\ref{fig10} (a), (d), and (g), which are nothing but the bulk NLS. By making the superlattices, the nodal lines change from a loop [Figs.~\ref{fig10} (a), (d), and (g)] into multiple loops [Figs.~\ref{fig10} (b), (c), (e), (f), (h), and (i)]. Some loops run across the entire Brillouin zone. These results are in good agreement with the results of the continuum model in Sec.~\ref{subsec2-3}.

\subsection{\label{subsec4-4}Superlattice with magnetization}

Similarly to Sec.~\ref{sec3}, we call the case of the magnetization ${\bm m}$ parallel to the stacking direction ${\bm n}$ pattern B-1 and that of the magnetization ${\bm m}$ perpendicular to the stacking direction ${\bm n}$ pattern B-2.
\subsubsection{\label{subsubsec4-4-1}Superlattice Pattern B-1}

\begin{figure}[]
\centering
\includegraphics[width=8.5cm,height=6.0cm]{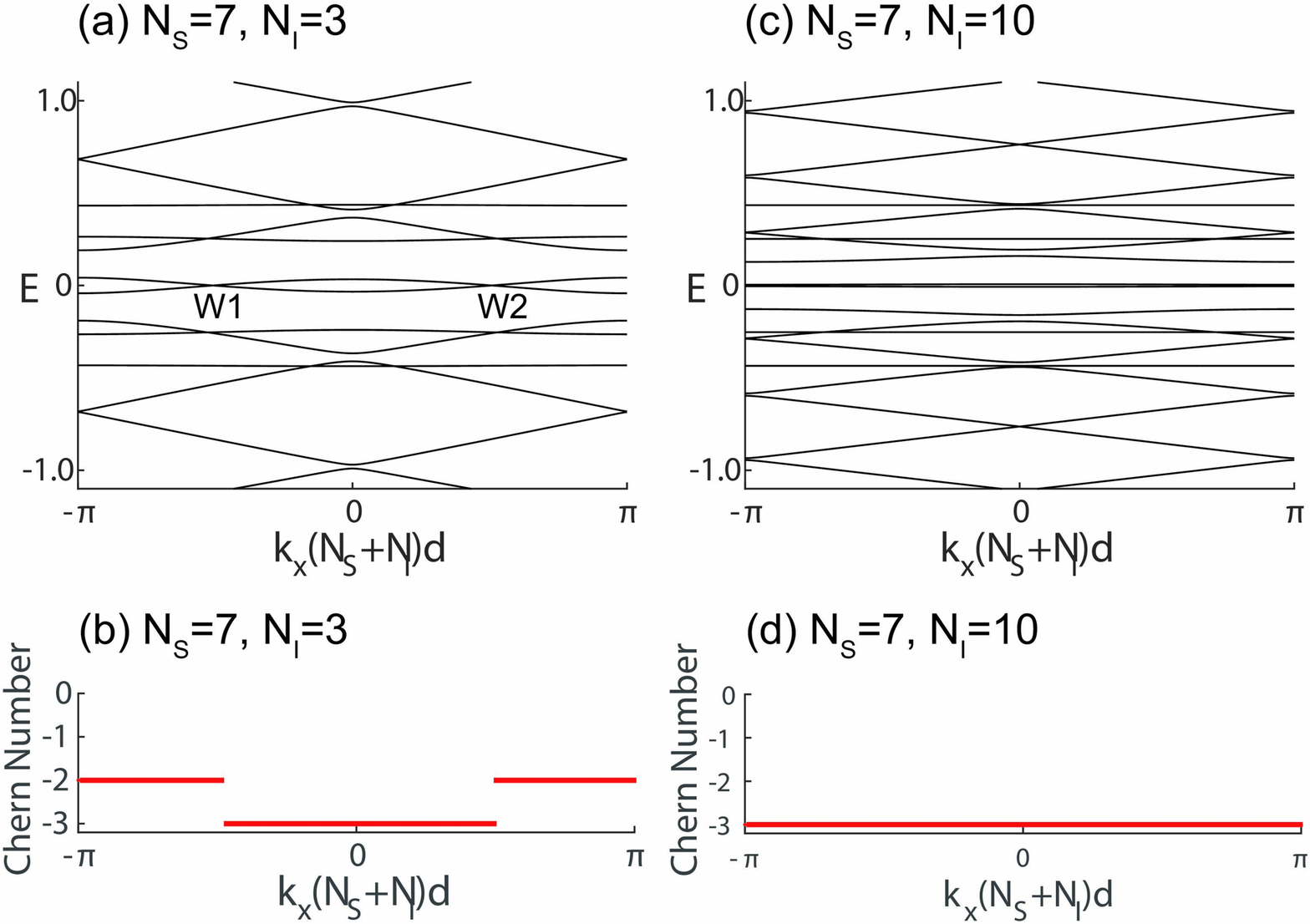}
\caption{\label{fig11} Band structure for $k_y=k_z=0$ and distribution of the Chern number in the superlattice with the magnetization in the stacking direction. (a) and (b) shows the case with $N_S=7$ and $N_I=3$, where the superlattice is in the WSM phase. The monopole ${\rm W}_1$ and the anti-monopole ${\rm W}_2$ appear in the bulk. (c) and (d) shows the case with $N_S=7$ and $N_I=10$, where the superlattice is in the QAH phase. We set the parameters with $V_S=0.5$, $V_I=0.1$, $d=2$, $g_\perp=2$, $g_z=1.2$, $v=1$, and $m=0.5$.}
\end{figure}

To investigate the effect of the magnetization, we add a perturbation term $V_T=\displaystyle\frac{m}{d}\sin k_yd~\sigma_x$ to the Hamiltonian Eq.~(\ref{eq8}). As aforementioned, this term represents the magnetization along the stacking direction. As shown in Ref.~\onlinecite{Okugawa2017}, the point-node degeneracies (Weyl nodes) survive if there are intersections between the $C_2$-invariant axis and the nodal lines, and the superlattice becomes the WSM. For example, when $N_S=7$ and $N_I=3$ shown in Fig.~\ref{fig10} (h), by introducing the magnetization, the $C_2$-invariant axis along the $k_x$ axis crosses the circular nodal line. In this case, we indeed numerically confirm that the superlattice becomes the WSM phase as shown in Figs.~\ref{fig11} (a) and (b). On the other hand, the magnetization opens the band gap if there are no intersections. When $N_S=7$ and $N_I=10$ as shown in Figs.~\ref{fig11} (c) and (d), the superlattice becomes the QAH phase with the Chern number $-3$. This Chern number corresponds to the number of the pair of the almost straight nodal lines. This result agrees with that of the effective model in Sec.~\ref{subsec3-1}.

\subsubsection{\label{subsubsec4-4-2}Superlattice Pattern B-2}

\begin{figure}[]
\centering
\includegraphics[width=8.5cm,height=3.8cm]{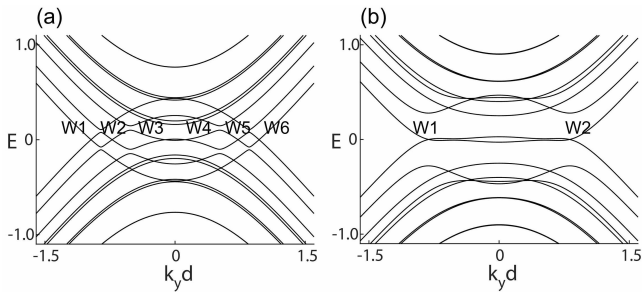}
\caption{\label{fig12} Band structure for $k_x=k_z=0$ in the superlattice with the magnetization in the $k_y$ axis, when $N_S=7$, $N_I=10$. We set the parameters as $V_S=0.5$, $V_I=0.1$, $d=2$, $g_\perp=2$, $g_z=1.2$, and $v=1$. The value of the parameter $m$ is $m=0.1$ in (a) and $m=1$ in (b). In (a), three pairs of the Weyl nodes appear in the bulk, while only one pair of the Weyl nodes survives by increasing the magnetization.}
\end{figure}

Next, we study the superlattice with the magnetization in the $k_y$ axis, which is perpendicular to the stacking direction. To do this, we add the perturbation term $V_T^\prime=\displaystyle\frac{m}{d}\sin k_xd~\sigma_x$ to the Hamiltonian Eq.~(\ref{eq8}). In Sec.~\ref{subsec3-2}, we find that there exist more than one pairs of the Weyl nodes in this case. In the lattice model, we numerically confirm that it is indeed the case as shown in Fig.~\ref{fig12} (a). While there are more than one pairs of the Weyl nodes when the magnetization is sufficiently small, only a single pair of the Weyl nodes survives by increasing the magnetization [Fig.~\ref{fig12} (b)]. Namely, the superlattice eventually behaves similarly to the WSM superlattice with pattern A in Ref.~\onlinecite{Yokomizo2017} by increasing the magnetization. 

\section{\label{sec5}Summary and Discussion}

In this paper, we studied properties of the NLS superlattices using the effective model and the lattice model. First of all, we investigated the superlattices with the stacking direction being either perpendicular or parallel to the plane where the nodal line appears, and  called them pattern A and pattern B, respectively.  In the superlattice with pattern A, we found a phase transition from the NLS phase to the NI phase by increasing the thickness of the NI layer. On the other hand, in pattern B, we showed that single nodal line is folded in the superlattice because of the periodicity of the Brillouin zone and this nodal line change into multiple nodal lines by increasing the thickness of the NI layer. Some nodal lines run across the entire Brillouin zone and they gradually become almost straight. Nodal lines always exist since the superlattice has both the IS and the TRS.

Furthermore, we introduced magnetization into the NLS-NI superlattices, and studied their properties. We then compared the results with our previous work on the WSM-NI superlattices \cite{Yokomizo2017}, because the NLS phase becomes the WSM phase by adding magnetization \cite{Okugawa2017}. We studied the two cases, with the magnetization ${\bm m}$ being perpendicular and with ${\bm m}$ parallel to the stacking direction ${\bm n}$.  In particular, the QAH phase can appear in the superlattice similarly to the WSM superlattice when ${\bm m}$ is parallel to ${\bm n}$. To realize the QAH phase, we need both a gap and a band inversion. First, the magnetization can open the gap when there are no intersections between the $C_2$-invariant axis and the nodal lines \cite{Okugawa2017}. Second, by increasing the thickness of the NLS layer, the Brillouin zone becomes narrower, and the nodal lines are folded at the Brillouin zone boundary. When ${\bm m}$ is added, it gives rise to a movement of the Weyl nodes along the $C_2$-invariant axis across the whole Brillouin zone, giving rise to the band inversion. This band inversion occurs multiple times by increasing the thickness of the NLS layer or the radius of the nodal line. Here, we propose that the NLS superlattice with the magnetization can realize the QAH phase with large Chern number, which has been a long-standing issue in this field \cite{Jiang2012,Yokomizo2017,Wang2013,Fang2014,Skirlo2014,Skirlo2015,Perez2015}.

On the other hand, when ${\bm m}$ is perpendicular to ${\bm n}$, some pairs of the Weyl nodes appear in the NLS-NI superlattice with the magnetization. This is in contrast with the result in the WSM-NI superlattice \cite{Yokomizo2017}, where the WSM superlattice has only a single pair of the Weyl nodes. Namely, although these two cases can be regarded as WSM-NI superlattices, the number of pairs of Weyl nodes in the superlattice is different between them. This difference comes from the bulk-band structure of the constituent WSM. In the WSM \cite{Okugawa2014} used in Ref.~\onlinecite{Yokomizo2017}, the bulk gap is wide except for the Weyl nodes. On the other hand, in our model of the NLS with the magnetization, the gap is very narrow along the original nodal line. Therefore, multiple pairs of the Weyl nodes can appear in the superlattice along the original nodal line with the narrow gap. Thus, the distribution of the gap size in the ${\bm k}$-space in the WSM determines how the Weyl nodes appear in the WSM-NI superlattice.

In this our work, we have focused on the NLS with the type-I dispersion. Recently, in addition to the nodal line, the NLSs with type-I\hspace{-.1em}I, hybrid \cite{Heikkila2015,Hyart2016,Li2017}, quadratic, and cubic \cite{Yu2018} dispersion have attracted much attention. We expect that the superlattice composed of the NLS with type-I\hspace{-.1em}I or hybrid dispersion behaves similarly with that of type-I dispersion because it is described by a model similar to Eq.~(\ref{eq1}), with an additional term proportional to the unit matrix. On the other hand, the NLS with quadratic or cubic dispersion proposed in Ref.~\onlinecite{Yu2018} is quite different from these with linear dispersion studied in the present paper. The quadratic or cubic nodal lines \cite{Yu2018} come from space-group symmetry and are confined on high-symmetry lines. Therefore, as long as the superlattice does not break the original space-group symmetry, the nodal lines will persist.

In recent years, considerable effort has been devoted to realize the topological phases in superlattices. For example, in Ref.~\onlinecite{Xu2015}, the authors show by first-principle calculation that a superlattice of two trivial ferromagnetic insulator shows a QAH phase. In Ref.~\onlinecite{Gaoyuan2018}, a multilayer of a ferromagnetically doped TI and a NI results in a QAH phase with a large Chern number. However, experimental realizations of the topological phases in the NLS superlattice have not been reported thus far, to the authorsf knowledge. It is interesting to investigate how these superlattices can be realized in future works.
\begin{acknowledgments}
This work was supported by Grant-in-Aid for Scientific Research (Grants No. 26287062 and No. 18H03678) by MEXT, Japan, by CREST, JST (Grant No. JPMJCR14F1), and by MEXT Elements Strategy Initiative to Form Core Research Center (TIES). Kazuki Yokomizo also was supported by International Research Center for Nanoscience and Quantum Physics, Tokyo Institute of Technology, and JSPS KAKENHI (Grant No. 18J22113).
\end{acknowledgments}
\appendix
\section{\label{sec6}Quantum well of a NLS}

Physically, the limit $b\rightarrow\infty$ in the NLS superlattice corresponds to a quantum well of the NLS with the well width $a$. We study this quantum well in this appendix. Here, the ${\bm k}$-space becomes two-dimensional and expressed as $\left(k_y,k_z\right)$ perpendicular to the well direction. By continuity of the wave function, we can obtain the relation between the energy eigenvalues $E$ and the well width $a$ as

\begin{equation}
\tan\sqrt{V_S\pm E-k_y^2}a=\frac{2\sqrt{V_S\pm E- k_y^2}\sqrt{V_I\mp E+ k_y^2}}{-2k_y^2+V_S-V_I\pm2E}.
\label{eq10}
\end{equation}

\begin{figure}[ht]
\centering
\includegraphics[width=9.0cm,height=3.4cm]{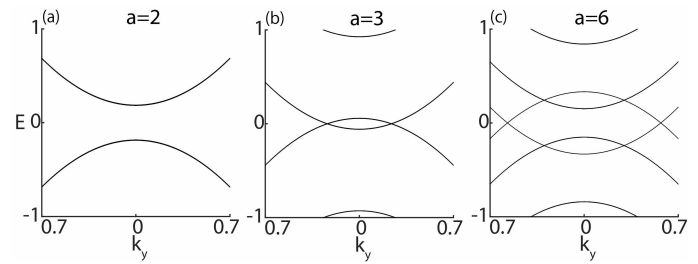}
\caption{\label{fig13}Energy bands of the quantum well of the NLS with the well width (a) $a$=2, (b) $a$=3, and (c) $a$=6. We set the value of the parameters as $V_S=0,5$ and $V_I=1$.}
\end{figure}

\noindent To find the gapless states, we set $k_z=0$ since they appear on the $C_2$-invariant axis, the $k_y$ axis. The energy bands are shown in Fig.~\ref{fig13}. The gapless states are given by $E=0$. We note that in the limit $b\rightarrow\infty$, the almost straight nodal lines in Figs.~\ref{fig4} (b) and (c) becomes the gapless states in Fig.~\ref{fig13} (b) and (c), respectively .

%
\nocite{*}

\providecommand{\noopsort}[1]{}\providecommand{\singleletter}[1]{#1}%

\end{document}